
\input harvmac
\input tables
\input epsf
\Title{hep-th/9511065}
{\vbox{\centerline{Tests for C-theorems in 4D}}}
\bigskip
\centerline{F. Bastianelli\footnote{$^*$}{email:
bastianelli@imoax1.unimo.it}}
\vglue .5cm
\centerline{Dipartimento di Fisica,
Universit\`a di Modena,}
\centerline{and}
\centerline{INFN, Sezione di Bologna, Italy.}

\vskip 2cm

\noindent

A proof for a non-perturbative $C$-theorem in four dimensions,
capturing the irreversibility of the renormalization group flow
in the space of unitary quantum field theories, has not been
accomplished, yet. We test the conjectured $C$-theorems
using the exact results recently obtained in $N=1$ supersymmetric
gauge theories. We find that the flow towards the infrared region
is consistent with the main proposals for a $C$-theorem.

\Date{11/95}
\def\sq{\,\raise.5pt\hbox{$\mbox{.09}{.09}$}\,}

Zamolodchikov's $C$-theorem
\ref\Z{ A.B. Zamolodchikov, JETP Lett. 43 (1986) 730;
Sov. J. Nucl. Phys. 46 (1987) 1090.}
is a standard tool in the studies of
two dimensional quantum field theories. It  states that
there exists a function $C(g^i)$ of the couplings $g^i$ which
is decreasing along the renormalization group trajectories
(leading to the infrared), and which is stationary at the fixed points,
where it coincides with the central charge of the corresponding
conformal field theories. The theorem makes precise the idea that
the detailed information on the short distance degrees of freedom
is lost in the renormalization group flow.
Extending the theorem to four dimensional field theories has been harder
then expected. It is a task that has not been fully accomplished, yet.
On the contrary, there are hints that such a theorem may not be valid
in four dimensions (e.g. see the comments in ref. \ref\F{
A.H. Castro Neto and E. Fradkin, cond-mat/9302009,
Nucl. Phys. B400 (1993) 525.}).
Nevertheless, many important ingredients which are believed to enter
the proof of a  4D $C$-theorem have been carefully discussed
\ref\C{ J.L. Cardy, Phys. Lett. B215 (1988) 749.}\ref\O {H. Osborn,
Phys. Lett. B222 (1989) 97\semi
I. Jack and H. Osborn, Nucl. Phys. B343 (1990) 647.}\ref\CFL{ A.
Cappelli, D. Friedan and J.I. Latorre, Nucl. Phys. B352
(1991) 616.}\ref\Sh{G.
Shore, Phys. Lett. B253 (1991) 380; B256 (1991) 407.}\ref\CLV{A.
Cappelli, J.I. Latorre and X. Vilas\'\i s-Cardona, hep-th/9109041,
Nucl. Phys. B376 (1992) 510.}.
The interest for such a 4D $C$-theorem would be considerable.
For example, it would allow to test
non-perturbative phenomena like confinement, chiral symmetry breaking,
supersymmetry breaking and the Higgs mechanism. On the other hand,
exacts results on $N=1$ supersymmetric gauge
theories exhibiting a plethora of such non-perturbative phenomena
have been recently obtained by Seiberg and collaborators
\ref\Su{N. Seiberg, hep-th/9402044, Phys. Rev. D49 (1994) 6857.}\ref\Sd{N.
Seiberg, hep-th/9411149, Nucl. Phys. B435 (1995) 129.}\ref\SI{K.
Intriligator and N. Seiberg, hep-th/9509066, and references therein.}.
This progress has shed some light
on the otherwise poorly understood subjects of
renormalization group flows, fixed points and
conformal field theories in four dimensions.
It seems therefore natural to test the conjectured $C$-theorems
against these exact results.
It is the purpose of this note to carry out such tests.
These are the first non-perturbative and non-trivial tests
which are possible in four dimensions.
We have found no counterexample to the conjectured
$C$-theorems. It is possible that a proof for the case of
supersymmetric theories may be easier,
if indeed such a theorem exists.

To start with we recall that in the renormalization
group flow the ultraviolet (UV) and the infrared (IR)
fixed points are described by conformal field theories.
These theories are characterized by the coefficients appearing
in the trace anomaly that arises once the theory is put on a
curved background. The trace anomaly can be parametrized as follows
\eqn\ta{ \langle \Theta \rangle = a F - b G + c \sq R, }
where $\langle \Theta \rangle $ is the one-point
function of the trace of the stress tensor
in a gravitational background, $F$ is the square of
the Weyl conformal tensor, $G$ is the topological Euler density and
$R$ is the scalar curvature.
For free theories the coefficients $a,b$ and $c$ have been computed,
since in this case a one-loop calculation
suffices \ref\cd{S. Christiensen and M. Duff,
Nucl. Phys. B170 (1980) 480\semi
Birrell and Davies \lq\lq Quantum Fields in Curved Space\rq\rq
(Cambridge U. Press, Cambridge, 1982).}.
Note that $c$ has not a universal meaning, and its value
can be changed at will by adding to the effective action
a local counterterm  proportional to the integral of
$R^2$. It is a coefficient that depends on the renormalization
scheme chosen.
The universal coefficients are instead
\eqn\coeff{\eqalign{
a &= {1\over{1920 \pi^2}} (N_0 + 3 N_{1\over 2} +
12 N_1)\cr
b &= {1\over{11520 \pi^2}} (2 N_0 + 11 N_{1\over 2} + 124 N_1), \cr}}
where $N_0, N_{1\over 2}, N_1$ are the number of real spin-$0$, real (Majorana)
spin-${1\over 2}$ and real spin-$1$ fields, respectively.

Now let us recall the main proposals for a $C$-theorem in 4D. Cardy's
suggestion \C\ is to consider the one-point function of the stress tensor
integrated over a sphere $S$ of constant radius
\eqn\one{ C = \kappa \int_S \langle \Theta \rangle { \sqrt g } d^4 x,}
where $\kappa$ is a numerical factor that depends on how one chooses
to normalize the $C$-function.
At the fixed point it reduces to the coefficient $b$ appearing in
the trace anomaly.
A related proposal due to Osborn \O\ is to take as $C$-function the
coefficient $b$ in the trace anomaly even off criticality,
modified suitably to make it satisfy an equation
similar to the one appearing in Zamolodchikov's work.
Also this function reduces
to the coefficient $b$ of the trace anomaly
when approaching the critical points.
The problem with these proposals is that it has not been possible
to prove that the candidate
$C$-functions are monotonically decreasing in the
infrared. Nevertheless, no counterexample has been found, yet.
Another proposal due to Cappelli, Friedan and Latorre \CFL\
identifies a $C$-function
by using a spectral representation of the
stress tensor and constructing a
reduced spectral density for the spin-$0$ intermediate states.
In this approach unitarity is manifest,
and it is used to prove  monotonicity.
However, it has not been shown if such a function has a definite
meaning at the fixed point.  It has only been checked that for
free spin-$0$ and spin-$1\over2$ fields it
coincides with the coefficient $a$ of the trace anomaly. In the
following we will assume this to be true in general, and test this
assumption. It is interesting to note that both coefficients
$a$ and $b$ should be positive, as they are supposed to measure
the number of degrees of freedom. The fact that $a$ is positive has
been proved in refs. \ref\CD{D. Casper and M. Duff,
Nucl. Phys. B82 (1974) 147.} and \CFL, while positivity of $b$ is for the
time being an empirical observation.

Let us now consider certain $N=1$ supersymmetric
gauge theories in which the UV and IR fixed point structures are
under control \Su\Sd\SI. The prototype is $SU(N_c)$ with $N_f$
scalar superfields in the fundamental and $N_f$ scalar superfields
in the anti-fundamental. The one-loop beta function is
\eqn\oneloop{ \beta(g) = - {g^3 \over{16 \pi^2}}\biggl [
{3\over 2}I({\rm Adj})-{\epsilon \over 2}N_f I({\rm Fund})\biggr ],}
where $I({\rm Adj}) = 2 N_c$ and $I({\rm Fund})= 1$
are the indices in the adjoint
and in the fundamental representations,
respectively, and $\epsilon=2$ to take into account that the fundamental
representation is complex.
Asymptotic freedom is achieved for $N_f < 3 N_c$. In this range the UV
fixed point is a free theory of $(N_c^2 -1)$
vector multiplets and $2 N_f N_c$ scalar multiplets.
If the fixed point is a free theory, the $C$-function reduces to
the sum of the central charges carried by the free fields.
We will denote these central charges by $C_S$ and $C_V$
for the scalar and vector multiplets, respectively.
We can take this central charges to be,
up to an irrelevant normalization factor present in eq. \coeff,
\eqn\central{\eqalign{ & a_S = 5, \ \ \ \   a_V = 15 ,\cr
& b_S = 15, \ \ \ \ b_V = 135, \cr}}
according to whether the $C$-function reduces at the critical point to the
$a$ or $b$ coefficient of the trace anomaly.
These values are easily obtained recalling that a scalar
multiplet contains two real spin-$0$ fields and a Majorana
fermion, while a vector multiplet contains a massless spin-$1$
field and a Majorana fermion.
Thus, at the UV fixed point the $C$-function is given by
\eqn\uno{C_{UV} = (N^2_c -1) C_V + 2 N_f N_c C_S .}
The IR structure of this theory has been analyzed in refs. \Su\Sd.
First of all, for $N_f=0$
one has a pure super Yang-Mills theory which confines and develops
a mass gap. Thus, the IR is described by the trivial theory
containing only the vacuum state, and it has $C_{IR}=0$.
The conjectured $C$-theorems are obviously satisfied.
For $ 0 < N_f < N_c$ there exists no vacuum.
For $ N_f = N_c$ there exists a smooth
quantum moduli space  described by the
vacuum expectation values of $N_f^2$ meson $M_i^j$,
one baryon $B$ and one antibaryon $\bar B$ scalar superfields
constrained by the equation
$ \det M - B \bar B = \Lambda^{2 N_c}$.
The massless fluctuations are those satisfying the constraint:
there are $(N_f^2 + 2 -1)$ free massless scalar superfields, and
they contribute
\eqn\due{ C_{IR} = (N_f^2 + 1 ) C_S. }
It is immediate to verify that $ \Delta C \equiv C_{UV} - C_{IR}>0 $.
For $N_f = N_c +1 $ the quantum moduli space contains singularities
associated with additional massless particles.
The maximum number of massless particles is at the
origin of the moduli space.
The corresponding IR fixed point is described by
$N_f^2$ free massless mesons and $2 N_f$
 free massless baryons, contributing
\eqn\tre{ C_{IR} = (N_f^2 + 2 N_f ) C_S. }
Once again $ \Delta C >0 $.
In the range $ N_c+2 \le N_f \le {3\over 2} N_c$,
present for $N_c\ge 4$, the IR description seems problematic.
However, Seiberg has proposed a dual
description  using \lq\lq magnetic" variables
in terms of which the theory becomes free (free magnetic phase).
This magnetic theory is described  by an $SU(N_f - N_c)$
gauge field,   $N_f$ types of quarks, $N_f$ types of
antiquarks and $N_f^2$ mesons.
All these fields are free in the IR and contribute
\eqn\quattro{ C_{IR} =  [(N_f - N_c)^2 -1 ] C_V +
[2 N_f ( N_f - N_c) + N_f^2] C_S. }
One can check that $\Delta C >0$. Note that
the more stringent test is achieved at $N_f = {3\over 2}N_c$,
where $\Delta C = {3\over 4} N_c^2 (C_V - C_S) $.
In the range ${3\over 2} N_c < N_f < 3 N_c$ the IR is described by a
superconformal field theory which can be parametrized using either
the electric or the magnetic variables.
In both variables the theory is interacting, and we cannot trust
one-loop results for the trace anomaly.
It would be very interesting to find a way of computing
exactly the trace anomaly for these interacting superconformal
field theories.
In the range $N_f \ge 3 N_c$ the theory ceases to be asymptotic
free. We collect these results in table 1
and plot some of these values in figures 1 and 2
for the reader's convenience.

\textfont0=\sevenrm
\scriptfont0=\fiverm
\textfont1=\seveni
\scriptfont1=\fivei
\textfont2=\sevensy
\scriptfont2=\fivesy
\thicksize=1pt
\vskip12pt
\begintable
\tstrut    $ N_f$ | $C_{UV}$ | $C_{IR}$ | $\Delta C $
\crthick   $0$   | $(N_c^2\!\!-\!\!1) C_V$ | $0$ |
 $(N_c^2\!\!-\!\!1) C_V$
\cr
$0  \! < \!\! N_f \!\! < \!\! N_c $ | $ (N_c^2 \!\! - \!\! 1) C_V
\!+\! 2 N_c N_f C_S $ | $?$  | $?$
\cr
$ N_c $ | $(N_c^2\!\!-\!\!1) C_V \!+\! 2 N_c^2 C_S $|
$(N_c^2\!+\!1)C_S$ |
$(N_c^2 \!\!-\!\!1) (C_V\! +\! C_S)$
\cr
$ N_c \!+\!1 $ | $(N_c^2\!\!-\!\!1) C_V \!+\! 2 N_c (N_c\!+\!1) C_S $|
$(N_c\!+\!1)(N_c\!+\!3)C_S$ |
$(N_c^2 \!\!-\!\!1) C_V \!+\! (N_c\!+\!1)(N_c\!\!-\!\!3) C_S$
\cr
$ N_c \!+\! 2 $ | $(N_c^2\!\!-\!\!1) C_V\!+\!2 N_c (N_c\!+\!2) C_S $|
$ 3 C_V \!+\! (N_c\!+\!2)(N_c\!+\!6)C_S$  |
$(N_c^2 \!\!-\!\!4) C_V\! +\! (N_c\!+\!2)(N_c\!\!-\!\!6) C_S$
\cr
$N_c\!+\!2 \! < \!\!N_f \!\!< \!\!{3\over 2}N_c $
| $ (N_c^2\!\!-\!\!1) C_V \!+\!
2 N_c N_f C_S $ |
$[(N_f\!\!-\!\!N_c)^2\!\!-\!\!1] C_V\!+\!(3N_f\!\!-\!\! 2 N_c)N_f C_S$|
$ (2N_c \!\!-\!\! N_f)N_f C_V \!+ \!  (4N_c \!\!-\!\! 3N_f)N_f C_S $
\cr
$ {3\over 2}N_c $ | $(N_c^2\!\!-\!\!1) C_V \!+\! 3 N_c^2 C_S $ |
$ ({1\over 4}N_c^2 \!\!-\!\!1 )C_V \!+\! {15 \over4} N_c^2C_S$  |
${3\over 4} N_c^2  (C_V \!\!-\!\! C_S)$
\cr
$ {3\over 2}N_c \!< \!\! N_f \!\!< \!  3N_c $ |
$(N_c^2\!\!-\!\!1) C_V \!+ \!
2 N_c N_f C_S $ | $ ? $| $?$
\endtable
\textfont0=\tenrm
\scriptfont0=\sevenrm
\textfont1=\teni
\scriptfont1=\seveni
\textfont2=\tensy
\scriptfont2=\sevensy
\medskip
\noindent {\bf Table 1:} Values of $C_{UV}$, $C_{IR}$ and $ \Delta C$
for $SU(N_c)$ with $N_f$ flavours. The question mark indicates that
the value of the $C$-function is unknown.

The previous pattern generalizes to other gauge groups with minor
theory dependent modifications.
The case of $SO(N_c)$ with $N_f$ flavours in the vector
representation is interesting since it shows many new non-perturbative
phenomena \ref\SIdue{K. Intriligator and N. Seiberg, hep-th/9503179,
Nucl. Phys. B444 (1995) 125.}.
The UV region is asymptotically free for $N_f < 3 (N_c-2)$,
and it has $C_{UV}= {1\over 2}N_c (N_c-1) C_V + N_f N_c C_S.$
In the IR for $0<N_f < N_c-4$ the theory has no vacuum state
because of the dynamical generation of
a superpotential. For $N_f=N_c-4$ there are two inequivalent phase
branches: one with a dynamically generated superpotential and without
a vacuum state, the other with a smooth moduli space of physically
inequivalent vacua. On this second branch the theory confines
without chiral symmetry breaking, and the massless mesons contribute
$C_{IR}= {1\over 2} N_f (N_f +1) C_S.$
For $N_f = N_c-3$ again we find two inequivalent phase branches: one
without a vacuum state, the other with a moduli space
containing singularities. These singularities describe
extra massless states on top of the mesons.
The most singular point is at the origin,
where the quantum theory has $N_f$ extra massless
composite scalar fields, and, correspondingly,
$C_{IR}= [N_f + {1\over 2} N_f (N_f+1) ] C_S.$
For $N_f=N_c-2$ the theory gets into an abelian
Coulomb phase (i.e. there is a massless photon)
with a moduli space containing singularities associated
to massless monopoles and dyons. The singularity associated to
the highest number of massless monopoles, and therefore to the
highest value of the $C$-function, gives
$C_{IR}= C_V + [2N_f + {1\over 2} N_f (N_f +1) ] C_S$.
For $N_c-1 \le N_f \le {3\over2}(N_c-2)$ there is a free
magnetic phase described by an $SO(N_f-N_c+4)$ gauge theory with $N_f$
flavours in the fundamental plus ${1\over2}N_f (N_f+1)$
gauge singlet mesons, giving
$C_{IR} = {1\over2} (N_f -N_c +4)(N_f-N_c+3) C_V + [N_f(N_f-N_c+4)
+{1\over2} N_f (N_f+1) ] C_S.$
Increasing further $N_f$ the theory gets into an interacting
non-abelian phase, and eventually looses ultraviolet
asymptotic freedom.
One can easily check that $\Delta C >0$ in all the above cases.

The $Sp(2N_c)$ gauge theory with $2N_f$ flavours of matter
fields \ref\IP{K. Intriligator and P. Pouliot, hep-th/9505006,
Phys. Lett. B353 (1995) 471.}
is asymptotically free for $N_f < 3 (N_c+1)$, and it has
$ C_{UV} = N_c (2 N_c +1) C_V + 4 N_c N_f C_S .$
The IR is as follows. For $0 < N_f \le N_c$ there is no vacuum. For
$N_f = N_c+1$ there exists a smooth quantum moduli space with
massless scalar excitations contributing
$C_{IR} = [N_f(2N_f-1) - 1] C_S.$
Also for $N_f= N_c +2$ there are only massless mesons.
At the origin of the moduli space they give
$C_{IR} = N_f(2N_f-1) C_S.$
For $N_c+3 \le N_f \le {3\over 2} (N_c +1)$ the theory is in  a free
magnetic phase described by an $Sp(2N_c-2N_f-4)$ gauge field
with $2N_f$ flavours of matter in the fundamental and $N_f (2N_f -1)$
free mesons, implying
$C_{IR} = (N_f-N_c -2) (2N_f -2N_c -3) C_V + [2 N_f (2N_f - 2N_c -4 )
+ N_f(2N_f -1)] C_S .$
In all these cases the $C$-theorems are satisfied.
Increasing further the number of flavours,
the theory enters in a phase described
by an interacting superconformal theory, and for $N_f \ge 3 (N_c+1)$
ultraviolet asymptotic freedom is lost.

The exceptional groups have not been
explicitly treated in the literature, yet, except for
the case of $G_2$ with $N_f$ flavours in the
fundamental \ref\igor{ I. Pesando, hep-th/9506139, Mod. Phys.
Lett. A10 (1995) 1871.}\ref\Gid{S.B. Giddings and J.M. Pierre,
hep-th/9506196.}\ref\Pouli{P. Pouliot, hep-th/9507018,
Phys. Lett. B359 (1995) 108.}.
In this model asymptotic freedom is achieved for
$N_f < 12$, where $C_{UV} = 14 C_V + 7N_f C_S$.
In the IR the interesting cases are for
$N_f=4$, where $C_{IR}= 14 C_S$, and $N_f=5$, where at the
origin of the moduli space $C_{IR} = 30 C_S$.
Clearly $\Delta C > 0$ in both cases.
There is no region with a free magnetic phase.

This completes the analysis of models with matter fields
transforming in the fundamental representation of simple
gauge groups. More complicate models
are those with matter fields in other representations. An example
is the $SU(N_c)$ gauge theory with $N_f$ flavours in the fundamental,
one flavor in the antisymmetric representation and $N_c+N_f-4$
flavors in the anti-fundamental ($N_c>2$ otherwise the antisymmetric
representation reduces to the singlet).
This model is interesting since it
has chiral gauge couplings and can break supersymmetry.
It has been recently discussed in
refs. \ref\Ber{M. Berkooz, hep-th/9505067.}\ref\PT{E. Poppitz
and S.P. Trivedi, hep-th/9507169.}\ref\PP{P. Pouliot, hep-th/9510148.}.
It has asymptotic freedom for $N_f < 2 N_c +3$, and in this range
$C_{UV}= (N_c^2-1)C_V + {1\over2} (4N_f + 3N_c -9)N_c C_S$.
In the IR for $N_f=3$ there is a smooth quantum moduli space with
a set of mesons and baryons, giving
$C_{IR}= {1\over2} (N_c^2+3N_c+2) C_S$.
For $N_f=4$ at the origin of moduli space  we find instead
$C_{IR}= {1\over2} (N_c^2 +7 N_c +18)C_S$. In both cases $\Delta C>0$.
Dual descriptions have been proposed in refs. \Ber\ and \PP, however
it is not clear to us if there exists a free magnetic phase
allowing further tests.

An even more complicated model which can be used to test the
$C$-theorem is that introduced by Kutasov \ref\Kut{D.
Kutasov, hep-th/9503086, Phys. Lett. B351 (1995) 230.}.
It requires a superpotential to identify a dual description.
Kutasov's model consists of an $SU(N_c)$ gauge theory with
$N_f$ flavours in the fundamental, $N_f$ flavours in the
anti-fundamental and a field $X$ in the adjoint.
The theory contains also a superpotential $W \sim X^3$.
It is asymptotically free for
$N_f<2N_c$. Kutasov has showed that
in the range ${1\over 2} N_c \le N_f \le {2\over 3}N_c$
the theory flows to an IR fixed point
which can be described by a free magnetic $SU(2N_f-N_c)$
gauge theory with $N_f$ fields in the fundamental, $N_f$ fields
in the anti-fundamental, a
field in the adjoint and $2N_f^2$ gauge singlets.
Thus one can compute
\eqn\a{\eqalign{
C_{UV}&= (N_c^2-1) C_V + (2N_fN_c +N_c^2-1)C_S \cr
C_{IR}&= [(2N_f -N_c)^2-1) C_V +
[2N_f(2N_f -N_c) +(2N_f -N_c)^2-1 + 2N_f^2]C_S, \cr}}
and check that $\Delta C>0$ in the allowed range of $N_f$.
This model has been generalized in
refs. \ref\ASY{O. Aharony, J. Sonnenschein and S.
Yankielowicz, hep-th/9504113, Nucl. Phys. B449 (1995)
509.}\ref\KS{D. Kutasov and A. Schwimmer,
hep-th/9505004, Phys. Lett. B354 (1995)
315.}\ref\KSS{ D. Kutasov, A. Schwimmer and
N. Seiberg, hep-th/9510222.}
by using more complicated superpotentials
containing higher powers of the field $X$.
These superpotentials allow for a richer pattern of IR fixed
points and corresponding dual descriptions.
However, the extra terms in the superpotentials
are non-renormalizable. They prevent the identification of
a UV fixed point, and we have not been able to use
such theories for our tests (put differently, the theories
with the more general superpotentials should be considered only as
low-energy effective descriptions).

Certainly there are many more models present in the literature
that could be used for our purposes, and we have not been able to
be exhaustive. Nevertheless, we have not found
a single counterexample. It is interesting to note that
whenever the $C$-theorem is on the verge of breaking down,
also the free field description of the IR fixed point breaks down,
saving the various conjectures.
This was, for example, the case of $SU(N_c)$ with $N_f={3\over2}N_c$
flavors, which is the value giving the upper
limit of the free magnetic phase.
If the free magnetic phase would continue beyond that limit,
the $C$-theorems would break down at
$N_f= {2C_V+4C_S\over{C_V+3C_S}}N_c$.
Of course,  this is not the case since
${2C_V+4C_S\over{C_V +3C_S}} > {3\over 2}$.
Supersymmetry and duality have been very useful for checking the
conjectured $C$-theorems.
Supersymmetry constrains the structure of the low-energy effective
action, and determines many of its properties.
Duality is essential to study at weak coupling
the long distance properties of some of the models considered.
As we have seen the conjectured $C$-theorems survive a wide set
of non-trivial tests.
Hopefully, supersymmetry could be used to find a simplified
proof for a $C$-theorem.

\vskip 1cm
\centerline{{\bf Acknowledgments}}
I wish to thank Andrea Cappelli for interesting discussions
and suggestions.
\vfill
\eject

\midinsert
\hfil
\epsfbox{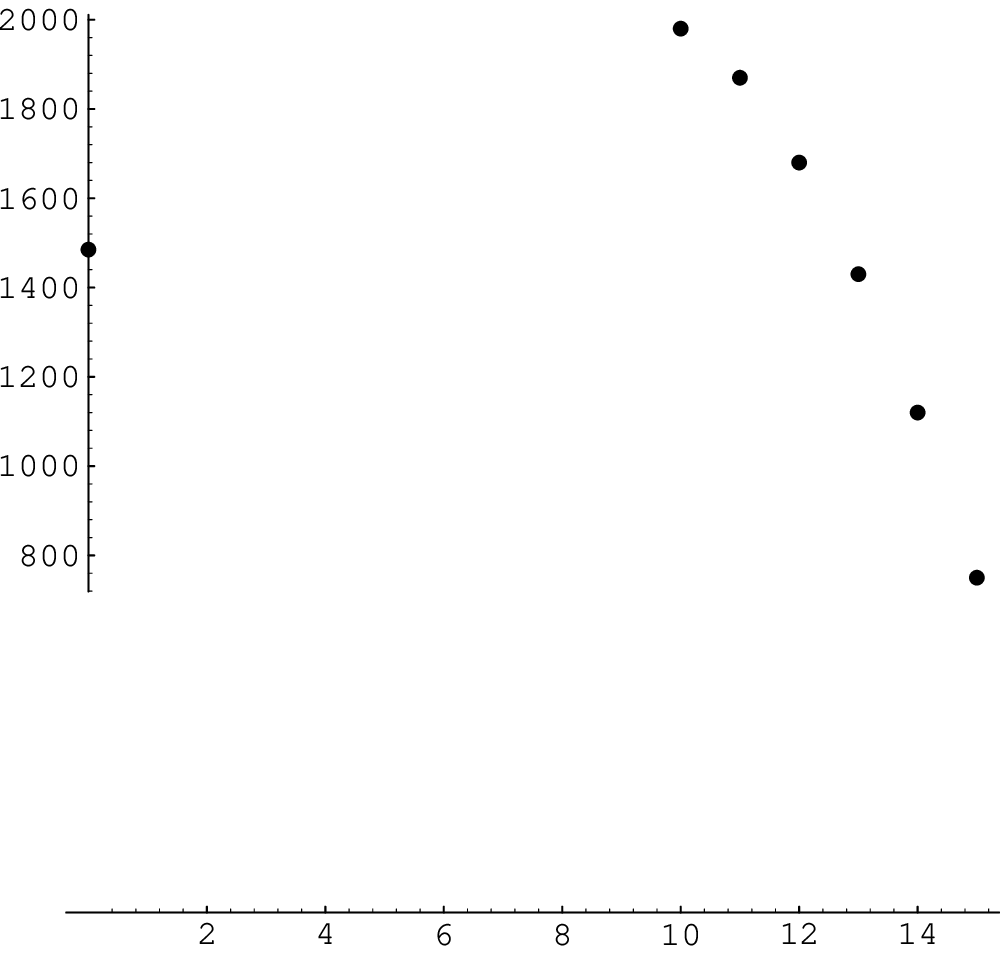}
\hfil
\endinsert
\noindent {\bf Fig.~1.} Plot of $\Delta C $ versus $N_f$ for $SU(10)$.
For definiteness we have used the $a$ coefficient of the trace anomaly
as fixed point value of the $C$-function. The plotted values can
be computed from table 1.
\vfill \eject

\midinsert
\hfil
\epsfbox{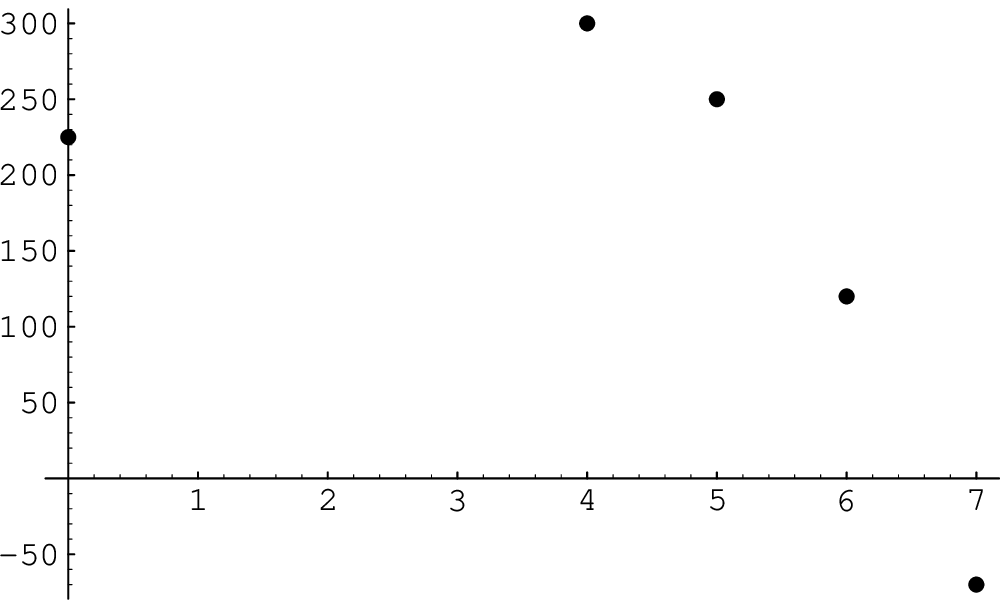}
\hfil
\endinsert
\vskip 2mm
\noindent {\bf Fig.~2.} Plot of $\Delta C$ versus $N_f$ for $SU(4)$.
The value at $N_f=7$ is out of the IR free magnetic phase, but we
have computed it by assuming that the free magnetic phase would
still hold. In that case the $C$-theorem related to the $a$
coefficient would break down since $\Delta C$ becomes negative.
\listrefs
\end